\documentstyle[aps,prd,preprint,eqsecnum,epsf]{revtex}

\newcommand{\be}{\begin{equation}}
\newcommand{\ee}{\end{equation}}
\newcommand{\ba}{\begin{eqnarray}}
\newcommand{\ea}{\end{eqnarray}}
\newcommand{\bas}{\begin{eqnarray*}}
\newcommand{\eas}{\end{eqnarray*}}
\newcommand{\hsp}{\hspace{.5cm}}

\newcommand{\spaceand}{\hsp {\rm and} \hsp}

\tightenlines

\begin{document}

\draft

\title{Nut Charge, Anti-de Sitter Space and Entropy}
\author{S.W. Hawking\thanks{email:  S.W.Hawking@damtp.cam.ac.uk}, 
        C.J. Hunter\thanks{email:  C.J.Hunter@damtp.cam.ac.uk} and } 
\address{Department of Applied Mathematics and
      Theoretical Physics, University of Cambridge,
      \\Silver Street, Cambridge CB3 9EW, United Kingdom }
\author{Don N. Page\thanks{email:  don@phys.ualberta.ca}} 
\address{CIAR Cosmology Program, Theoretical Physics Institute, Department of
Physics, \\
University of Alberta, Edmonton, Alberta, Canada, T6G 2J1}
\date{4 September 1998}

\maketitle

\begin{abstract}
It has been proposed that spacetimes with a $U(1)$ isometry group 
have contributions to the entropy from Misner strings as well as 
from the area of $d-2$ dimensional fixed point sets. In this paper we test this 
proposal by constructing Taub-Nut-AdS and Taub-Bolt-AdS solutions 
which are examples of a new class of asymptotically locally anti-de
Sitter spaces. We find that with the additional contribution from 
the Misner strings, we exactly 
reproduce the entropy calculated from the action by the usual thermodynamic
relations. This entropy has the right parameter dependence to agree 
with the entropy of a conformal field theory on the boundary, which is 
a squashed three-sphere, at least in the limit of large squashing. 
However the conformal field theory and the normalisation of the entropy 
remain to be determined. 
\end{abstract}


\narrowtext

\section{Introduction}
\label{sec:Intro}

It has been known for quite some time that black holes have entropy. The 
entropy is 
\be
 S = \frac{{\cal A}}{4G}, 
\ee
where $\cal A$ is the area of the horizon and $G$ is Newton's constant.   In any
dimension $d$, this
formula holds for black holes or black branes that have 
a horizon, which is a $d - 2$ dimensional fixed point set of a $U(1)$ isometry
group. However it has recently been shown \cite{HawkingHunter98}
that entropy can be
associated with a more general class of spacetimes. In these metrics, the 
$U(1)$ isometry group
can have fixed points on surfaces of any even co-dimension, and the spacetime
need not be asymptotically flat or asymptotically anti-de Sitter. In this more
general class, the entropy is not just a quarter the area of the $d-2$
dimensional fixed point set.

 Among the more general class of spacetimes for which entropy
can be defined, an interesting case is those with nut charge. Nut charge can be
defined in four dimensions \cite{GibbonsHawking79}
 and can be regarded as a magnetic type of mass.
Solutions with nut charge are not asymptotically flat (AF)
in the usual sense. Instead, they are said to be asymptotically locally flat
(ALF). In the Euclidean regime, 
in which we shall be working, the difference can 
be described as follows. An AF metric, like Euclidean
Schwarzschild, has a boundary at infinity that is an $S^2$ of radius $r$
times an $S^1$, whose radius is asymptotically constant. To get finite values
for the action and Hamiltonian, one subtracts the values for 
periodically identified flat space.  In ALF metrics, on the other
hand, the boundary at infinity is an $S^1$ bundle over $S^2$. These bundles are
labeled by their first Chern number, which is proportional to the nut charge. If
the first Chern number is zero, the boundary is the product $S^2 \times  S^1$, 
and the metric is AF. However, if the first Chern number is $k$, then the
boundary is a squashed $S^3$ with $|k|$ points identified around the $S^1$
fibers. Such ALF metrics cannot be matched to flat space at infinity to give a
finite action and Hamiltonian, despite
a number of papers that claim it can be done. The best that one can do is match
to the self-dual multi-Taub-NUT solutions \cite{Hawking77}. 
These can be regarded as defining 
the vacuums for ALF metrics.

In the self-dual Taub-NUT solution, the $U(1)$ isometry
group has a zero-dimensional fixed point set at the center, called a nut.
However, the same ALF boundary conditions admit another Euclidean solution,
called the Taub-Bolt metric \cite{Page78}, in which the nut is replaced by a 
two-dimensional
bolt. The interesting feature is that, according to the new definition of
entropy, the entropy of Taub-Bolt is not equal to a quarter the area of the
bolt, in Planck units. The reason is that there is a contribution to the entropy
from the Misner string, the gravitational counterpart to a Dirac string for a
gauge field.

The fact that black hole entropy is proportional to the area of
the horizon has led people to try and identify the microstates with states on
the horizon. After years of failure, success seemed to come in 1996, with the 
paper of Strominger and Vafa \cite{StromingerVafa96},
which connected the entropy of certain black holes with a system of D-branes.
With hindsight, this can now be seen as an example of a duality between a
gravitational theory in asymptotically anti-de Sitter space, and a conformal
field theory on its boundary. It would be interesting if similar dualities could
be found for solutions with nut charge, so that one could verify that the
contribution of the Misner string was present in the entropy of a
conformal field theory. This would be particularly significant for solutions 
like 
Taub-Bolt, which don't have a spin structure. It would show that the duality 
between
anti-de Sitter space and conformal field theories on its boundary did not
depend on supersymmetry or string theory.

In this paper, we will describe the
progress we have made towards establishing such a duality. We have found a
family of Taub-Bolt anti-de Sitter solutions. These Euclidean metrics
are characterized by an integer $k$, and a positive real parameter, $s$. The
boundary at large distances is an $S^1$ bundle over $S^2$, with first Chern 
number $k$. If $k=0$, the boundary is a product, $S^1 \times S^2$, and the 
space is asymptotically anti-de Sitter, in the usual sense. 
But if $k$ is not zero, the metrics
are what may be called asymptotically locally anti-de Sitter, or ALAdS. The
boundary is a squashed $S^3$, with $k$ points identified around the $U(1)$
direction. This is just like ALF metrics. But
unlike the ALF case, the squashing of the $S^3$ tends to a finite limit
as one approaches infinity. This means that the boundary has a well defined
conformal structure. One can then ask whether the partition function and
entropy of a conformal field theory on the boundary is related to the action
and entropy of these ALAdS solutions.

To make
this question well posed we have to specify the reference backgrounds with
respect to which the actions and Hamiltonians are defined. 
Like in the ALF case, a squashed $S^3$ cannot be imbedded in Euclidean anti-de 
Sitter. Therefore one cannot use it as a
reference background to regularize the action and Hamiltonian. Instead, one has
to use Taub-NUT anti-de Sitter, which is a limiting case of our family. If $|k|$
is greater than one, there is an orbifold singularity in the reference
backgrounds, but not in the Taub-Bolt anti-de Sitter solutions. These orbifold
singularities in the backgrounds could be resolved by replacing a small
neighbourhood of the nut by an ALE metric. We shall therefore take it that the 
orbifold singularities are harmless.

Another issue that has to be resolved is what
conformal field theory to use on the squashed $S^3$. Here we are on
shakier ground. For five-dimensional anti-de Sitter space, there are good
reasons to believe that the boundary theory is large $N$ Yang Mills. But on the
three-dimensional boundaries of four-dimensional anti-de Sitter spaces, 
Yang Mills theory is not conformally invariant. The best that 
we can do is calculate
the determinants of free fields on the squashed $S^3$, and see if they have the
same dependence on the squashing as the action. Note that as the boundary is
odd dimensional, there is no conformal anomaly. The determinant of a conformally
invariant operator will just be a function of the squashing. We can then
interpret the squashing as the inverse temperature, and get the number of
degrees of freedom from a comparison with the entropy of ordinary black holes
in four-dimensional anti-de Sitter.  

\section{Entropy}

We now turn to the question of how one can define the entropy of a spacetime. 
A thermodynamic ensemble is a collection of
systems whose charges are constrained by Lagrange multipliers.  One such charge
is the energy or mass $M$, with the Lagrange multiplier being the inverse
temperature, $\beta$. But one can also constrain the angular momentum $J$,
and gauge charges $q_i$. The partition function for the ensemble is the 
sum over all states, 
\be
 {\cal Z} = \sum e^{-\mu_i K_i},
\ee 
where $\mu_i$ is the Lagrange multiplier associated with the charge $K_i$. 
Thus, it can also be written as 
\be
 {\cal Z} = {\rm Tr}\, e^{-Q}.
\ee
Here $Q$ is the operator that generates a Euclidean time translation
$\Delta \tau = \beta$, a rotation $\Delta \phi = \beta \Omega$ and a gauge
transformation $\alpha_i = \beta \Phi_i$, where $\Omega$ is the angular
velocity and $\Phi_i$ is the gauge potential for $q_i$. 
In other words, $Q$ is
the Hamiltonian operator for a lapse that is $\beta$ at infinity, a shift 
that
is a rotation through $\Delta \phi$, 
and gauge rotations $\alpha_i$.  This means that the partition function
can be
represented by a Euclidean path integral over all metrics which are periodic
at infinity under the combination of a Euclidean time translation by $\beta$, a
rotation through $\Delta \phi$, and a gauge rotation $ \alpha_i$. 
The lowest order
contributions to the path integral for the partition function will come from
Euclidean solutions with a $U(1)$ isometry that agree with the periodic 
boundary conditions at infinity.

The Hamiltonian in general relativity or supergravity
can be written as a volume integral over a surface of constant $\tau$, 
plus surface integrals over its boundaries.  The notation used will be that
of \cite{HawkingHunter98}.  The volume integral is
\be
 H_c = \int_{\Sigma_\tau} d^{d-1}x\,\left[N{\cal H} + N^i{\cal H}_i +
       A_0(D_i E^i - \rho) + \sum_{A=1}^M \lambda^A C^A \right],
\ee
and vanishes by the constraint equations. 
Thus the numerical value of the Hamiltonian comes
entirely from the surface terms, 
\be
 \label{eqn:Hamiltonian_b}
 H_b = -\frac{1}{8\pi G} \int_{B_\tau} \sqrt{\sigma}[ Nk +
         u_i(K^{ij}-Kh^{ij})N_j + 2 A_0 F^{0i} u_i + f(N,N^i,h_{ij},\phi^A) ].
\ee
The action can be related to the Hamiltonian
in the usual way, 
\begin{equation}
 I = \int d\tau \left[ \int_{\Sigma_\tau} d^{d-1}x (P^{ij}\dot{h}_{ij} +
          E^i\dot{A}_i + \sum_{A=1}^N \pi^A \dot{\phi}^A ) + H\right].
\end{equation}
Because the metric has a $U(1)$ isometry all dotted 
quantities vanish. Thus
\be
I =\beta H.
\ee

If the solution can
be foliated by a family of surfaces that agree with Euclidean time at infinity,
the only surface terms will be at infinity.  In this case, a solution can be
identified under any time translation, rotation, or gauge transformation at
infinity. This means that the action will be linear in $\beta$, $\Delta \phi$, 
and $\alpha_i$,
\be
I =\beta H_{\infty} = \beta M + (\Delta \phi) J +\alpha_i q_i.
\ee
If one takes such a linear action to be $(-\log {\cal Z})$, and applies the 
standard thermodynamic relations, one finds the entropy is zero.

The situation is very
different, however, if the solution cannot be foliated by surfaces of constant 
$\tau$, where $\tau$ is the parameter of the $U(1)$
isometry group that agrees with the periodic identification at infinity. The
breakdown of foliation can occur in two ways. The first is at fixed points of
the $U(1)$ isometry group. These occur on surfaces of even co-dimension. 
Fixed point sets of co-dimension two play a special role.  We shall refer to 
them as bolts.  Examples include the horizons of non-extreme black holes and 
p-branes, but there can be more complicated cases, as in Taub-Bolt.

The other way the foliation by
surfaces of constant $\tau$ can break down is if there are what are called 
Misner strings. To explain what they are, we write the metric in the 
Kaluza-Klein form with respect to the $U(1)$ isometry group,
\begin{equation} 
 ds^2 = \exp\left[-\frac{4\sigma}{\sqrt{d-2}}\right](d\tau + \omega_i dx^i)^2 +
     \exp\left[\frac{4\sigma}{(d-3)\sqrt{d-2}}\right] \gamma_{ij}dx^idx^j.
\end{equation}
The one-form, $\omega_i$, the dilaton, $\sigma$,
and the metric, $\gamma_{ij}$, can be regarded as fields on $\Xi$, the space of 
orbits of
the isometry group. If $\Xi$ has homology in dimension two, 
the Kaluza-Klein field strength $F$ can have non-zero integrals 
over two-cycles. This means that the one-form, $\omega_i$, will have 
Dirac strings in $\Xi$. In turn, this will mean that the foliation of the 
spacetime $\cal M$ by surfaces of constant $\tau$ will break down on 
surfaces of co-dimension two, called Misner strings.

In order to do a Hamiltonian treatment using surfaces of constant $\tau$,
one has to cut out small neighbourhoods of the fixed point sets and the Misner
strings. This modifies the treatment in two ways.  First, the surfaces of
constant $\tau$ now have boundaries at the fixed point sets and Misner 
strings, as well as the usual boundary at infinity. This means there can be 
additional surface
terms in the Hamiltonian. In fact, the surface terms at the fixed point sets are
zero, because the shift and lapse vanish there. On the other hand, at a Misner
string the lapse vanishes, but the shift is non-zero. The Hamiltonian can
therefore have a surface term on the Misner string, which is the shift times a
component of the second fundamental form of the constant $\tau$ surfaces. The
total Hamiltonian will be 
\be
 H = H_\infty +  H_{\rm MS},
\ee
i.e., the sum of this Misner string Hamiltonian and the Hamiltonian surface 
term at infinity.
As before, the action will be $\beta H$.   However, this will be
the action of the spacetime with the neighbourhoods of the fixed point sets and
Misner strings removed. To get the action of the full spacetime, one has to put
back the neighbourhoods. When one does so, the surface term associated with the
Einstein-Hilbert action will give a contribution to the action of minus area
over $4G$, for both the bolts and Misner strings, that is,
\be
 \label{eqn:Ham_action}
 I = \beta H_\infty + \beta H_{\rm MS} -\frac{1}{4G}({\cal A}_{\rm bolt} + 
     {\cal A}_{\rm MS}).
\ee
Here $G$ is Newton's constant in 
the dimension one is considering. The surface terms around lower dimensional 
fixed point sets make no contribution to the action.

The action of the spacetime, $I$, will be the lowest order contribution to 
$(-\log {\cal Z})$. 
But 
\be
\log {\cal Z} = S - \beta H_\infty.
\ee
So the entropy is
\be
 \label{eqn:entropy}
 S  =  \frac{1}{4}({\cal A}_{\rm bolt} + {\cal A}_{\rm MS}) - (\Delta \psi) H_{\rm MS}.
\ee
In other words, the entropy is the amount by which the action is less
than the value, $\beta H_\infty$, that it would have if
the surfaces of constant $\tau$ foliated the spacetime.

The formula (\ref{eqn:entropy}) for the entropy applies in any dimension, and 
for any class of
boundary condition at infinity. In particular, we can apply it to ALF 
metrics in four dimensions that have nut charge. In this case, the reference
background is the self-dual Taub-NUT solution. The Taub-Bolt solution has the
same asymptotic behaviour, but with the zero-dimensional fixed point
replaced by a two-dimensional bolt. The area of the bolt is $12 \pi N^2$,
where $N$ is the nut charge. The area of the Misner string is $-12 \pi N^2$.
That is to say, the area of the Misner string in Taub-Bolt is infinite, but it
is less than the area of the Misner string in Taub-NUT, in a well defined sense.
The Hamiltonian on the Misner string is $-N/8$.  
Again the Misner string Hamiltonian
is infinite, but the difference from Taub-NUT is finite. 
And the period, $\beta$, is $8\pi N$. Thus the entropy is 
\be
 S = \pi N^2. 
\ee
Note that this is less than a quarter the area of the bolt, which would give 
$3 \pi N^2$. It is the effect
of the Misner string that reduces the entropy.

\section{Entropy of Taub-Bolt-AdS}

The Taub-NUT-AdS metric can be obtained as a special case of the complex 
metrics given in \cite{PagePope} (see also \cite{Kramer80}).  The line element is 
\ba
 \label{eqn:NUTmetric}
 ds^2 & = & b^2E \left[ 
        \frac{F(r)}{E(r^2-1)}(d\tau + E^{1/2}\cos\theta d\phi)^2
        + \frac{4(r^2-1)}{F(r)} dr^2 \right. \nonumber \\ 
 & &   \left.     + (r^2-1)(d\theta^2 + \sin^2\theta d\phi^2) 
                             \right],
\ea
where
\be
 F_N(r,E)  = Er^4 + (4-6E)r^2 +  (8E-8)r + 4-3E, 
\ee
$E$ is an arbitrary constant which parameterizes the squashing, 
$b^2=-3/4\Lambda$, and $\Lambda<0$ is the cosmological 
constant.  The Euclidean time coordinate, $\tau$, has period is 
$\beta = 4\pi E^{1/2}$ and has a nut at $r=1$, which is the origin of the 
$\psi-r$ plane.
Asymptotically, the metric is ALAdS since the boundary is
a squashed $S^3$, rather than $S^1 \times S^2$.  

We can obtain another family of metrics from \cite{PagePope} that have the 
same asymptotic behaviour.  They are the Taub-Bolt-AdS metrics, which have
the same form as (\ref{eqn:NUTmetric}) but the function $F(r)$ is 
\be
 F_B(r,s)  =  Er^4 + (4-6E)r^2 + \left[-Es^3+(6E-4)s+\frac{3E-4}{s}\right]r + 
     4-3E,
\ee
where
\be
 E = \frac{2ks-4}{3(s^2-1)},
\ee
$k$ is the Chern number of the $S^1$ bundle and $s$ is an arbitrary parameter. 
In order to avoid curvature singularities, we must take $s>1$, $s>2/k$
and $r>s$.  The periodicity of the imaginary time 
is $4\pi E^{1/2}/k$, and it has a bolt at $r=s$, with area
\be
 {\cal A}_{\rm bolt} = \frac{8}{3}b^2\pi (ks-2).  
\ee
The boundary at infinity is a squashed $S^3$ with $|k|$ points 
identified on the $S^1$ fibre.  

  The action calculation is a fairly trivial combination of the original 
Schwarzschild-AdS action calculation \cite{HawkingPage83} and the more recent
understanding of the actions of metrics with nut charge \cite{Hunter98}.  
As mentioned in section \ref{sec:Intro}, in order to regularize the action
and Hamiltonian calculations, we need to choose a reference background.  Since
Taub-Bolt-AdS cannot be imbedded in AdS, we cannot use this as a background.
However, we can use a suitably identified and scaled Taub-NUT-AdS as a 
reference background.  We need the periodicity of the imaginary time 
coordinates to agree. This
means that for a Taub-Bolt-AdS metric with parameters $(k,s)$  we must take
the orbifold obtained by identifying $k$ points on the $S^1$ as the reference
background, rather than just Taub-NUT-AdS.  This will have a conical 
singularity at the 
origin, however, as mentioned before, we can smooth it out in a simple way,
and hence we can just ignore it, and treat the space as non-singular.  We then
need to scale the background imaginary time by $E^{1/2}/\tilde{E}^{1/2}$ so that
both imaginary time 
coordinates have the same  periodicity, namely $\beta = 4\pi E^{1/2}/k$.   
Finally, we require that the induced metrics agree sufficiently
well on a hypersurface of constant radius $R$, as we take $R$ to infinity.
This yields equations for both the $S^1$ and the $S^2$ metric components,
\ba
 \frac{EF_B(r,s)}{r^2-1} & = & 
   \frac{\tilde{E} F_N(\tilde{r},\tilde{E})}{\tilde{r}^2-1} \spaceand \\
 E(r^2-1) & = & \tilde{E}(\tilde{r}^2-1).
\ea
To sufficient order, this has the solution $\tilde{E} = \eta E$ and 
$\tilde{r} = \lambda r$, where 
\be
 \eta = 1-\frac{2\rho}{R^3}, \hsp \lambda=1+\frac{\rho}{R^3} \spaceand 
 \rho = \frac{(s-1)^2[E(s-1)(s+3) + 4]}{2sE}.
\ee
Hence the matched background metric is
\ba
 \label{eqn:NUTback}
 ds^2 & = & b^2\eta E\left[ 
    \frac{F_N(\lambda r,\eta E)}{E(\lambda^2 r^2-1)}(d\psi + 
        E^{1/2}\cos\theta d\phi)^2
        + \frac{4(\lambda^2 r^2-1)}{F_N(\lambda r,\eta E)} \lambda^2 dr^2 
    \right.   \nonumber \\
   & &    \left.  + (\lambda^2 r^2-1)(d\theta^2 + \sin^2\theta d\phi^2) 
                             \right],
\ea
with the function
\be
 F_N(\lambda r, \eta E) = 
    E\eta \lambda^4 r^4 + (4-6E\eta)\lambda^2r^2 +  (8E\eta-8)\lambda r + 
      4-3E\eta.
\ee

\begin{figure}[h]
 \begin{center}
 \epsfxsize=.75\textwidth
 \epsfbox{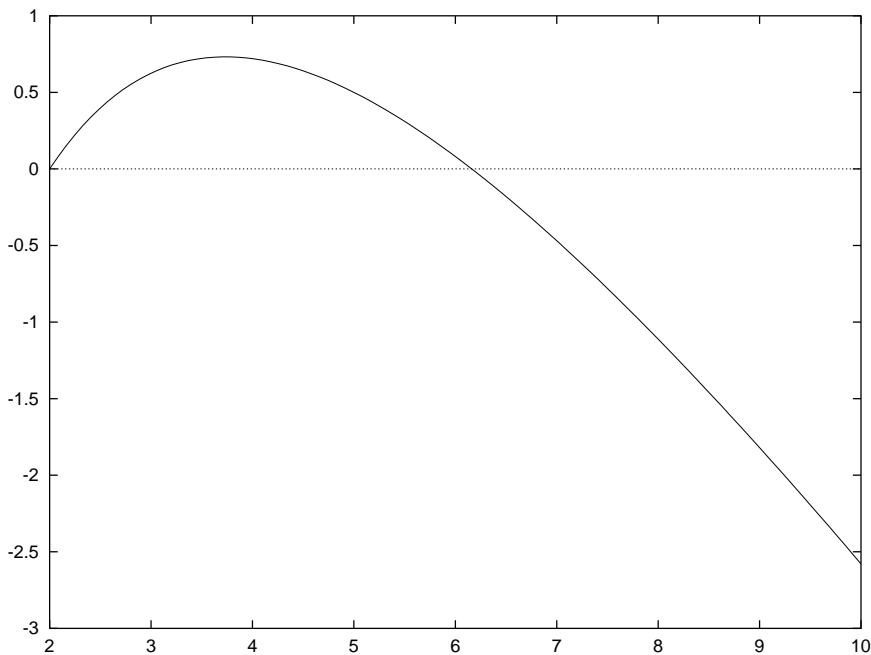}
 \end{center}
 \caption{The action $I$ as a function of $s$ for $k=1$ and $b^2=9/2\pi$, as
   given by equation (\ref{eqn:TBAdS_action}). The zero is at $s=3+\sqrt{10}$.}
 \label{fig:s_action}
\end{figure}

Calculating the action, we find that the surface terms cancel, just like in
the Schwarzschild-AdS case, so that the action
is given entirely by the difference in volumes of the metrics, 
\be
 \label{eqn:TBAdS_action}
 I  = -\frac{2\pi b^2}{9k} \frac{(ks-2)[k(s^2+2s+3) - 4(2s+1)]}{(s+1)^2}.
\ee
We see that the action will have zeros at up to 3 points,
\be
 s_\pm = \frac{4-k\pm\sqrt{16-4k-2k^2}}{k} \spaceand s_0 = \frac{2}{k}.
\ee
For the case $k=1$, there will only be one valid zero, $s_+=3+\sqrt{10}$.
The action will be positive for $s<s_+$, and negative for $s>s_+$.  When $k=2$,
all the zeros will coincide at the lowest value of $s=1$, and the action is
negative for any other value of $s$.  For larger values
of $k$, $s_\pm$ will be imaginary, $s_0<1$ and hence the action will always be
negative.  The action for $k=1$ is plotted in figure \ref{fig:s_action}.

The Hamiltonian calculation is more complicated than the simple action 
calculation completed above.
There will be two non-zero contributions to the Hamiltonian -- from the 
boundary at
infinity and from the boundary along the Misner string.  There is a third
boundary, around the bolt, but the Hamiltonian will vanish there.  Using the 
matched Taub-NUT-AdS metric from above, we find that
\be
 H_\infty = \frac{b^2}{9}\frac{(s-1)(ks-2)[k(s + 3) + 4]}{E^{1/2}(s+1)^2},
\ee
and
\be
 H_{\rm MS} = \frac{b^2}{3}\frac{(k-2s)(ks-2)}{E^{1/2}(s+1)^2}.
\ee
The area of the Misner string is larger in the background, and hence the net
area is negative,
\be
 {\cal A}_{\rm MS} = -\frac{32\pi b^2}{3} \frac{ks-2}{s+1},
\ee
while the area of the bolt is
\be
 {\cal A}_{\rm bolt} = \frac{8\pi b^2}{3}(ks-2).
\ee 
Substituting these values into the formula for the action 
(\ref{eqn:Ham_action}) we regain the expression (\ref{eqn:TBAdS_action}).
 
We are now in a position to use equation (\ref{eqn:entropy}) for the 
entropy.  We find that
\be
 \label{eqn:TB_entropy}
 S = \frac{2\pi b^2}{3k} \frac{(ks-2)[k(s^2+2s-1)-4]}{(s+1)^2}. 
\ee
Similar to the action, the entropy will have three possible zeros, 
\be
 s_\pm = \frac{-k\pm\sqrt{2k^2+4k}}{k}, \spaceand s_0 = \frac{2}{k}.
\ee
For $k=1$, all the zeros satisfy $s \leq 2$, while for $k=2$, the zeros 
are at $s \leq 1$.  Hence in these cases the entropy is never negative, and
is only zero at $(s=2,k=1)$ and $(s=1,k=2)$, which are exactly the
two points where the action vanishes.
For larger values of $k$, the zeros are all strictly less than 1, and hence
the entropy is always positive. 

One  can regard $\cal Z$
as the partition function  at a temperature 
\be
T =\beta ^{-1}=\frac{k}{4\pi E^{1/2}}.
\ee
If one then assumes that mass is the only charge that is constrained 
by a Lagrange multiplier (nut charge is fixed by the boundary conditions 
and hence does not need a Lagrange multiplier), then one 
can calculate the entropy from the standard thermodynamic relation 
\be
 S = \beta \frac{\partial I}{\partial \beta} - I = 
          2E\frac{\partial I}{\partial E} - I,
\ee
where we have made the approximation $I = -\log {\cal Z}$.  This yields the
same value as in (\ref{eqn:TB_entropy}) and so acts as a consistency
check on our formula for entropy. 

One can also calculate the energy, or mass of the system,
\be
 M = \frac{\partial I}{\partial \beta} = 
   \frac{b^2}{9}\frac{(s-1)(ks-2)[k(s + 3) + 4]}{E^{1/2}(s+1)^2} = H_\infty.
\ee
Again, this agrees with the Hamiltonian calculation. 

\begin{figure}[h]
 \begin{center}
 \epsfxsize=.75\textwidth
 \epsfbox{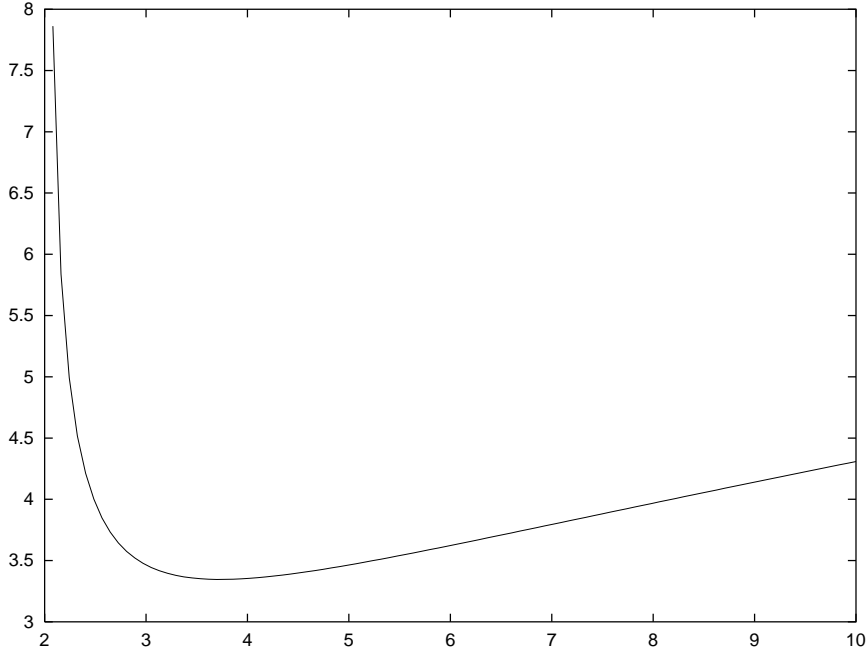}
 \end{center}
 \caption{The temperature $T=1/\sqrt{E}$ as a function of $s$ for $k=1$ and 
    $b^2=9/2\pi$.  The minimum value is at $s=2+\sqrt{3}$.}
 \label{fig:s_temp}
\end{figure}

Identical to the AdS case, there is a phase transition in the
ALAdS system (for $k=1$).  This can be seen by considering the behaviour of the
Taub-NUT-AdS and Taub-Bolt-AdS solutions as a function of temperature.  
There are no restrictions on the temperature of Taub-NUT-AdS, but, as can be
seen from figure \ref{fig:s_temp}, the temperature of Taub-Bolt-AdS has a
minimum value $T_0 = \sqrt{6+3\sqrt{3}}/(4\pi) \approx 0.836516303738/\pi$.
Hence, if we have $T<T_0$, the system will be in the Taub-NUT-AdS ground state.
As we increase $T$ above $T_0$, there are two possible
Taub-Bolt metrics with different mass values but the same temperature.  The one
with lower $s$ will be thermodynamically unstable, since it has negative
specific heat, $\partial M/\partial T$, while the
one with larger $s$ has positive specific heat, and hence will be stable.  The lower
$s$ branch has positive action, and hence will be less likely than the
background Taub-NUT-AdS.  The behaviour of the larger $s$ branch will depend on
$T$.
At temperatures below $T_1 = \sqrt{7+2\sqrt{10}}/(4\pi) \approx
0.912570384968/\pi$, the action will be
positive and the Taub-NUT-AdS background will be favoured.  But for $T$
greater than $T_1$, the negative action implies that the Taub-Bolt-AdS 
solution is preferred, and hence the Taub-NUT-AdS background will inevitably
decay into it. 

We can compare the local temperatures at the phase transition for the
Schwarzschild-AdS (k=0) and the Taub-Bolt-AdS ($k=1$ and
the degenerate case $k=2$) metrics.
In order to compare the temperatures in the different metrics, we want 
to rescale them so that the radii of the $S^2$ parts of their boundaries at 
infinity are one.
Hence, rescaling the $S^2\times S^1$ boundary of the 
Schwarzschild-AdS case
corresponds to multiplying the temperatures given in \cite{HawkingPage83} by
the quantity $b = \sqrt{-3/\Lambda}$ used in that paper, which is twice the 
$b$ used in our present paper.  In that case one gets $T^{k=0}_0 = \sqrt{3}/(2\pi)$ and 
$T^{k=0}_1 = 1/\pi$. 
In the Taub-Bolt-AdS case, the temperature at the
boundary with this rescaling is simply $(4\pi\sqrt{E})^{-1}$, as we have
defined it above.
The corresponding temperatures for the $k=1$ metric are 
$T^{k=1}_0 = \sqrt{2+\sqrt{3}}T^{k=0}_0/2 \approx 0.96593\, T_0^{k=0}$ and 
$T^{k=1} = \sqrt{7+2\sqrt{10}}/(4\pi) T^{k=0}_1\approx 0.91257\, T_1^{k=0}$
respectively.  For $k=2$, the minimum and critical temperatures coincide, and
they are $T^{k=2} = T_0^{k=0}/\sqrt{2} = \sqrt{3/8}T_1^{k=1}$.  The results are summarized in the table below: 
\begin{center}
 \begin{tabular}{l|l|l}
  \multicolumn{1}{c|} {$k$} & \multicolumn{1}{c|} {$\pi T_0$} & 
  \multicolumn{1}{c} {$\pi T_1$}  \\ \hline
  0 & 0.86660 & 1.0 \\
  1 & 0.83652 & 0.91257 \\
  2 & 0.61237 & 0.61237 
 \end{tabular}
\end{center} 
It is interesting that
the first two results are much closer together than they are to the $k=2$ value.

\section{Conformal Field Theory}

Formally at least, one can regard Euclidean conformal field theory 
on the squashed $S^3$ as a twisted $2+1$ theory on an $S^2$ of unit 
radius at a temperature $T=\beta^{-1}$. Thus, one would expect 
the  entropy to be proportional to $\beta^{-2}$
for small $\beta$. This dependence agrees with the expression that we have 
for the gravitational entropy of Taub-Bolt-AdS. To go further and 
obtain the normalisation and sub-leading dependence on $\beta$
would require a knowledge of the conformal field theory that we 
don't have. The best that we can do is calculate the determinants of 
conformally invariant free fields on the squashed $S^3$ 
and compare with the results for $S^2\times S^1$ and Schwarzschild-AdS. 
On $S^2\times S^1$ the determinants 
of conformally invariant free fields will be the same function of 
$\beta$, but this cannot be the case on the squashed $S^3$ 
because fermions have zero modes at an infinite number of values 
of the squashing, whereas a scalar field has a zero mode only at one 
value. Furthermore, Taub-Bolt-AdS solutions with $k$ odd do not have spin
structures. Thus if they are dual to a conformal field theory, it 
should be one without fermions. 

Similar work on Taub-NUT-AdS and Taub-Bolt-AdS for $k=1$ has been 
performed independently \cite{Chamblin98}.

\section{Acknowledgments}

CJH and DNP acknowledge the financial support of 
the Natural Sciences and Engineering Research Council of Canada.

\end{document}